\documentclass[superscriptaddress,twocolumn,10pt,prx,aps,showpacs,longbibliography,floatfix]{revtex4-2}
\usepackage{graphicx}
\usepackage[dvipsnames]{xcolor}
\usepackage{physics}
\usepackage{dcolumn}
\usepackage{bm}
\usepackage{comment}
\usepackage{siunitx}
\usepackage{mathtools}

\usepackage{float}
\usepackage{array}
\usepackage{amsbsy}
\usepackage[utf8x]{inputenc}
\usepackage{epstopdf}
\usepackage{amsmath,amssymb,amsfonts}
\usepackage{booktabs}
\usepackage{dsfont}
\usepackage{cprotect}
\usepackage{mathtools}
\usepackage[normalem]{ulem}
\usepackage{multirow}
\usepackage{verbatim}
\usepackage{tabularx}
\usepackage{enumitem}
\usepackage{url}
\usepackage{lipsum} 

\usepackage{mathtools}
\DeclarePairedDelimiter\ceil{\lceil}{\rceil}

\usepackage[colorlinks]{hyperref}
\hypersetup{
	plainpages=true,
	breaklinks=true,
	hypertexnames=false,
	pageanchor=true,
	colorlinks=true,
	linkcolor={blue},
	citecolor={red},
	urlcolor={blue},
	anchorcolor={black}
}

\newcommand{\LL}[0]{\mathcal{L}}

\newcommand{\D}[0]{\mathcal{D}}
\newcommand{\C}[0]{\mathcal{C}}

\renewcommand{\a}[0]{\hat{a}}
\renewcommand{\H}[0]{\hat{H}}
\newcommand{\J}[0]{\hat{J}}
\newcommand{\Gammao}[0]{\hat\Gamma}
\newcommand{\ad}[0]{{\hat{a}^\dagger}}
\newcommand{\sx}[0]{\hat\sigma^x}
\newcommand{\sy}[0]{\hat\sigma^y}
\newcommand{\sz}[0]{\hat\sigma^z}
\newcommand{\sm}[0]{\hat\sigma^-}

\newcommand{\rhot}[0]{\hat{\rho}(t)}
\newcommand{\rhoo}[0]{\hat{\rho}}
\newcommand{\sss}[0]{\hat{\rho}_{\rm ss}}

\newcommand{\B}[0]{\bm{B}}
\newcommand{\Bdot}[0]{\dot{\bm B}}
\renewcommand{\P}[0]{\bm P}
\newcommand{\btheta}[0]{\bm{\theta}}
\newcommand{\bthetadot}[0]{\dot{\bm{\theta}}}
\newcommand{\z}[0]{\bm{z}}
\newcommand{\Si}[0]{\bm{S}^{-1}}
\newcommand{\Bi}[0]{\bm{B}^{-1}}
\newcommand{\zdot}[0]{\dot{\bm z}}
\renewcommand{\S}[0]{\bm{S}}
\renewcommand{\L}[0]{\bm{L}}
\newcommand{\Ltil}[0]{\bm{\tilde{L}}}
\newcommand{\btau}[0]{\bm{\tau}}
\newcommand{\I}[0]{\mathds{1}}

\newsavebox{\mstrut}
\newcommand{\bbra}[1]{
    \sbox{\mstrut}{\(#1\)}
    \mathinner{\left\langle\kern-0.3\ht\mstrut\left\langle{#1}\right|\mkern-2mu\right|}
}
\newcommand{\kett}[1]{
    \sbox{\mstrut}{\(#1\)}
    \mathinner{\left|\mkern-2mu\left|{#1}\right\rangle\kern-0.3\ht\mstrut\right\rangle}
}
\newcommand{\kettbbra}[2]{
    \sbox{\mstrut}{\(#1\)}
    \mathinner{\left|\mkern-2mu\left|{#1}\right\rangle\kern-0.3\ht\mstrut\right\rangle\mkern-5.8mu\left\langle\kern-0.3\ht\mstrut\left\langle{#2}\right|\mkern-2mu\right|}
}

\usepackage[normalem]{ulem}
\renewcommand\sout{\bgroup\markoverwith{\textcolor{red}{\rule[0.5ex]{2pt}{0.8pt}}}\ULon}

\begin{document}

\author{Luca Gravina}
\email{luca.gravina@epfl.ch}
\affiliation{Institute of Physics, Ecole Polytechnique Fédérale de Lausanne (EPFL), CH-1015 Lausanne, Switzerland}
    \affiliation{Center for Quantum Science and Engineering, Ecole Polytechnique Fédérale de Lausanne (EPFL), CH-1015 Lausanne, Switzerland}
\author{Vincenzo Savona}
\email{vincenzo.savona@epfl.ch}
\affiliation{Institute of Physics, Ecole Polytechnique Fédérale de Lausanne (EPFL), CH-1015 Lausanne, Switzerland}
\affiliation{Center for Quantum Science and Engineering, Ecole Polytechnique Fédérale de Lausanne (EPFL), CH-1015 Lausanne, Switzerland}

\title{Adaptive variational low-rank dynamics for open quantum systems}

\begin{abstract}
We introduce a novel, model-independent method for the efficient simulation of low-entropy systems, whose dynamics can be accurately described with a limited number of states. 
Our method leverages the time-dependent variational principle to efficiently integrate the Lindblad master equation, dynamically identifying and modifying the low-rank basis over which we decompose the system's evolution. By dynamically adapting the dimension of this basis, and thus the rank of the density matrix, our method maintains optimal representation of the system state, offering a substantial computational advantage over existing adaptive low-rank schemes in terms of both computational time and memory requirements.
We demonstrate the efficacy of our method through extensive benchmarks on a variety of model systems, with a particular emphasis on multi-qubit bosonic codes, a promising candidate for fault-tolerant quantum hardware.
Our results highlight the method's versatility and efficiency, making it applicable to a wide range of systems characterized by arbitrary degrees of entanglement and moderate entropy throughout their dynamics. We provide an implementation of the method as a Julia package, making it readily available to use.
\end{abstract}

\date{\today}

\maketitle

\section{Introduction}

In the path towards scalable and reliable quantum hardware \cite{brink2018, cai2021, campbell2017, joshi2021, kempe2001, preskill2018, terhal2015, terhal2020}, the loss of quantum coherence due to interactions with an external environment presents a crucial limitation. While environmental interactions are often known for introducing noise and hindering quantum effects, under careful control, they can also serve as a resource, often harbouring surprising benefits not found in equilibrium systems. These features have long been exploited for various applications, including the preparation and stabilization of quantum states \cite{Verstraete_quantum_2009, polkovnikov_colloquium_2011, qisert_quantum_2015}, protocols for quantum error correction and suppression \cite{blume-kohout2010, kempe2001, knill2000, lidar2013, lidar2014, blume-kohout2010, gravina2023, lieu2023, lieu2020, leghtas2015, touzard2018}, enhancement of metrological protocols \cite{dicandia2021, fernandez-lorenzo2017, ilias2022}, and the stabilization and control of entangled, topological, and localized phases \cite{gong_topological_2018, kawabata_symmetry_2019, hamazaki_non-hermitian_2019, ippoliti_entanglement_2021, kawabata_entanglement_2023}.

Understanding the impact of both incoherent and coherent relaxation processes on quantum systems is therefore crucial for the progress of quantum technologies. Consequently, the development of fast and scalable methods for their classical simulation is equally important.
The exact numerical simulation of many-body quantum systems, however, suffers from the curse of dimensionality, namely the exponential growth of computational resources, scaling as $d^N$, required to simulate the dynamics of a system with $N$ identical modes, each with $d$ possible states. As a result, simulating even moderately large quantum systems becomes rapidly an intractable problem \cite{nori2014}.

When a system is not isolated, but couples to the degrees of freedom of an external environment, the two become entangled. The system's state is then represented by an Hermitian positive semi-definite operator, the density matrix, resulting from tracing out the environmental degrees of freedom. Under the assumption of a memoryless environment, the time evolution of the density matrix is governed by the Lindblad master equation \cite{gorini1978, lindblad1976, gorini1976} which has found application in a very broad range of physical settings \cite{rotter2015}. In this context, the curse of dimensionality becomes even more daunting, with computational requirements scaling as $d^{2N}$. 
Different approaches for mitigating the complexity of the task exist \cite{weimer2021}, each one compromising on the degree of accuracy with which different features of the statistical ensemble can be captured. Among these, Monte Carlo quantum trajectory methods \cite{daley2014, Verstraelen2023}, phase-space methods \cite{Polkovnikov2010, Deuar2021}, semiclassical approximations \cite{biella2016, Huybrechts2020}, tensor network techniques \cite{weimer2021, schollwck2005, schollwck2011, schollwck2019, eisert2013, Feldman2022, ors2014, cirac2021}, and variational approaches \cite{benjamin2019, endo2020, Reh2021, azad2023, wu2020, gerdts1997, raab1999, burghardt1999} possibly coupled with Monte Carlo sampling and neural network techniques \cite{carleo2019, nagy2019, Luo2020, Carrasquilla2019, yoshioka2019, vicentini2019, nagy2018}, stand out.

In recent years, another class of methods know as \emph{ensemble truncation methods} has emerged as a powerful tool for simulating isolated \cite{pagliantini2023, pagliantini2022, pagliantini2021, zoe2022, zoe2023} and dissipative quantum systems \cite{ciuti2021, chen2021, mccaul2021, rouchon2013}. This approach is based on the realization that many quantum systems, particularly those with low entropy, can be effectively represented by a density matrix of significantly lower rank than what the whole Hilbert space would require. This reduction is achieved by focusing on a subset of states that capture the essential structure of the statistical ensemble characterizing the mixed quantum state, thereby reducing computational complexity while maintaining accuracy.
The low-rank (LR) hypothesis applies to many systems of physical interest. Among these are systems weakly coupled to the environment, or systems initialized in a pure, i.e. zero-entropy state, in the early stage of their dynamics. Modern quantum-computing platforms, designed to minimize dissipation and noise, often fall in either of the above instances. 

A considerable challenge remains, however, determining how the LR states should evolve to optimally represent the density matrix along the time evolution, and in particular how the dimension of the LR subspace should vary to accommodate changes of the entropy over time. 
Building upon this premise, this paper introduces a novel method encapsulating the most relevant features of the ensemble truncation schemes within the robust framework of the time-dependent variational principle (TDVP) developed for open quantum systems in Refs.~\cite{weimer2015, doriol2014, doriol2015, benjamin2019}. Indeed, our method, which we deem \emph{LR-TDVP method}, effectively integrates the benefits of both approaches. On one side it leverages the dynamical truncation methods' ability to efficiently represent quantum states with minimal information loss and the dynamically adjustment of the basis' dimension to adapt to changes in the system's entropy. On the other it takes advantage of the variational principle to dynamically modify the LR basis states thus ensuring the optimal fidelity of the evolution. 
This combination is particularly suited for quantum systems characterized by an arbitrary degree of entanglement but moderate entropy as is often the case in modern quantum hardware. By addressing the challenges faced in the NISQ era, our method offers a significant advancement in the simulation of complex quantum systems, paving the way for new discoveries and applications in quantum computing and simulation.

Our method is designed to be universally applicable, independent of specific system characteristics such as spatial symmetries, particle statistics, geometry of the space, or topology of the interactions. To facilitate its use and integration into various research workflows, we have incorporated it as a method of the QuPhys library \cite{quphys}, a Julia-based \cite{bezanson2017julia} quantum physics toolkit. The library, along with our method, is readily accessible and can be found at the repository listed in \cite{git}.

\begin{figure*}[htb]
\center
% \hspace*{-1.em}
\includegraphics[width=0.94\textwidth]{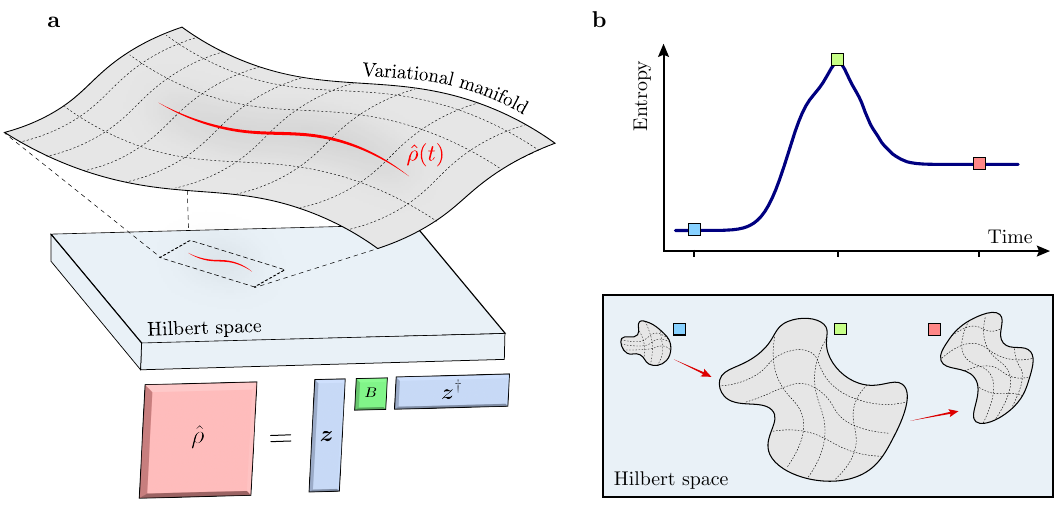}
\caption{\label{fig:artistic} 
Illustrative overview of the LR-TDVP Method. (a) Top: Pictorial representation of the density matrix $\rho(t)$ (depicted in red) at a generic time $t$ during the dynamics. The variational manifold selectively encompasses a specific region of the Hilbert space, capturing the predominant components of the statistical ensemble at time $t$. Bottom: Illustration of how memory efficiency is achieved by decomposing the $N_{\mathcal{H}}\times N_{\mathcal{H}}$ density matrix into smaller $N_{\mathcal{H}}\times M$ and $M\times M$ matrices. (b) Schematic representation of the dynamic adjustment of the variational manifold's dimension. Both the manifold and the rank of the truncated density matrix are dynamically modified to maintain a consistent level of accuracy, adapting to changes in the system's entropy. This adjustment reflects the principle that highly entropic systems necessitate a larger number of states for accurate representation, whereas low-entropy ensembles can be effectively described with fewer states.}
\end{figure*}

\section{The low-rank TDVP}
Let us consider an open quantum system whose dynamics is described by the Lindblad master equation 
\begin{equation}
\label{eqn:ME}
    \frac{\partial \rhoo}{\partial t} = \LL(\rhoo) = -i\comm*{\H}{\rhoo} + \sum_{\sigma=1}^D \D[\Gammao_\sigma]\rhoo ,
\end{equation}
where $\hat{\rho} \equiv \rhot$ (for brevity) is the system density matrix at time $t$, $\LL$ is the Liouvillian superoperator, and we take $\hbar=1$. While the coherent part of the dynamics is described by the Hamiltonian $\H$ acting on a Hilbert space $\mathcal{H}$ of dimension $N_{\mathcal{H}}$, the incoherent evolution is described by the dissipator 
\begin{equation}
    D[\Gammao] \hat{\rho} = \Gammao\hat{\rho}\, \Gammao^\dagger - \frac{1}{2}\acomm*{\Gammao^\dagger \Gammao}{\hat{\rho}},
\end{equation}
which formalizes the action of the jump operator $\Gammao$ on the system. 

The exact numerical simulation of this dynamics is generally computationally challenging due to the exponential scaling of the required resources, growing as the square of the Hilbert space dimension. 
The focus of our work is however on low-entropy systems, whose dynamics is known to be well captured, at all times, by a limited number of states $\{\ket{\varphi_k(t)}\,;\,k=1,\ldots,M\}$ with $M\ll N_{\mathcal{H}}$. These states span the \emph{low-rank} subspace $\mathcal{H}_{M} \subseteq \mathcal{H}$ which is assumed to encapsulate the essential structure of the statistical ensemble throughout the dynamics. 
In recent years, various heuristic algorithms have been developed that effectively truncate the density matrix $\rhoo$ to a rank-$M$ matrix, significantly smaller than what the whole Hilbert space would require. Such methods are collectively known as \emph{ensemble truncation methods} \cite{ciuti2021, mccaul2021, chen2021, rouchon2013}. Fundamental to these methods is their ability to dynamically adjust the truncation rank $M=M(t)$ at each time step, adapting to an entropy landscape that varies over time. However, a rigorous and systematic approach for selecting and adapting the low-rank basis states remains a challenge.

Concurrently, \emph{variational methods} have emerged as a versatile tool for circumventing the exponential space problem by considering trial states from a physically motivated, small subset of the exponentially large Hilbert space \cite{benjamin2019}.
In these methods, the density matrix $\rhoo(t) = \rhoo(\btheta(t))$ is expressed in terms of a set of variational parameters $\btheta(t)$ which evolve in time to guarantee, within the expressiveness of the ansatz, the optimal approximation of the evolution generated by the action of $\LL$ on $\rhoo(t)$.

In this work, we introduce the low-rank TDVP algorithm where the benefits of both classes of methods are united. By employing the McLachlan TDVP we rigorously estimate the optimal subspace $\mathcal{H}_{M(t)}$ over which to decompose the evolution at each time step (Sec.~\ref{sec:var_EOM}). 
Additionally, we develop schemes to dynamically adjust the dimension of this subspace in response to changes in the system's entropy (Sec.~\ref{sec:dynamical_rank}).
Deferring a detailed analysis of the ensemble truncation methods to App.~\ref{sec:DynCornerSpace}, we lastly compare our variational method to the heuristic algorithms established in \cite{ciuti2021, mccaul2021, chen2021} (Sec.~\ref{sec:comparison}). The main features of this method are summarized in Fig.~\ref{fig:artistic}.

\subsection{Variational equations of motion}
\label{sec:var_EOM}
The most general variational parametrization of an arbitrary statistical ensemble can be expressed as:
\begin{equation}
\label{eqn:LRansatz}
    \rhot = \sum_{i,j=1}^{M(t)} B_{ij}(t) \ketbra{\varphi_i(t)}{\varphi_j(t)},
\end{equation}
with $B_{ij}(t)$ being time-dependent, Hermitian coefficients [$B_{ij}(t) =
B_{ji}(t)^*$]. 
This formulation confines the dynamics within the variational manifold spanned by the states $\{\ket{\varphi_k(t)}\,;\,k=1,\ldots,M(t)\}$ whose dimension we adapt in time to accommodate changes in the system's entropy.
Throughout the paper we will refer to $\ket{\varphi_k(t)}$ as the \emph{variational} or \emph{low-rank states} and to $M(t)$ as the \emph{rank} of $\rhoo$.

Importantly, the variational method ensures a dynamical adjustment of the variational states, guaranteeing the optimal set of states is selected at all times to best approximate the evolution under $\LL$. To enable this dynamic adjustment, the variational states themselves must be parametrized.
In our approach, we adopt a full linear parametrization, decomposing $\ket{\varphi_k}$ on the computational basis $\{\ket{e_\alpha}\}_{\alpha=1,\ldots,N_{\mathcal{H}}}$ as:
\begin{equation}
\label{eqn:phi_LinearAnsatz}
    \ket{\varphi_k(t)} = \sum_{\alpha=1}^{N_{\mathcal{H}}} z_{\alpha,k}(t) \ket{e_\alpha}.
\end{equation}
For clarity, and where it does not lead to confusion, we will omit the time argument in subsequent discussions. While this work focuses on full linear parametrization, alternative parametrizations relying on tensor networks or neural network architectures are indeed possible and will be subject of future investigation.

The dynamical problem outlined in Eq.~\eqref{eqn:ME} can now be reformulated into a set of equations of motion (EOM) for the variational parameters, comprising the population matrix $\B=[B_{ij}]$ and the coefficient matrix $\z=[z_{\alpha k}]$. To derive these EOM, we apply the McLachlan variational principle, with the additional constraint of trace preservation:
\begin{equation}
\label{eqn:McLachlan}
\delta \left[ \norm{\left(\dv{t}-\LL\right)\rhoo\,}^2 + \lambda \dv{t}\Tr{\rhoo + \rhoo^\dagger} \right] = 0,
\end{equation}
where $\lambda$ is introduced as a Lagrange multiplier.

While Eq.~\eqref{eqn:McLachlan} is general in scope, substituting Eq.~\eqref{eqn:LRansatz} for $\rhoo$ yields the EOM specific for the chosen ansatz (the details of the derivation are provided in App.~\ref{App:EOM_derivation}):
\begin{equation}
\label{eqn:EOM}
    \begin{aligned}
        \Bdot &= \qty(\Si \L -\frac{\Tr{\Si\L}}M\I)\Si \\[0.2cm]
        \zdot &= \left(\Ltil- \z\Si\L\right)\Si\Bi,
    \end{aligned}
 \end{equation}
where $\Ltil = \LL(\rhoo)\z$, $\L = \z^\dagger\Ltil$, $\S=\z^\dagger\z$, while $\Si$ and $\Bi$ denote the inverse matrices of $\S$ and $\B$ respectively. Since both $\S$ and $\B$ may be singular, their inverse is often ill-defined. To overcome this problem, throughout the paper we adopt a smooth regularization criterion based on the singular value decomposition of the two matrices, as proposed in Ref.~\cite{medvidovic2023}. Details are deferred to App.~\ref{app:pseudoinverse}.

The dynamics of the system can now be obtained by numerical integration of Eq.~\eqref{eqn:EOM} which we perform using high-order adaptive-timestep solvers. Each time step requires the calculation of $\B^{-1}$ and of $\Ltil$. The first we perform using singular value decomposition, while the second only requires matrix multiplications, the most extensive being between (extremely sparse) $N_{\mathcal{H}} \times N_{\mathcal{H}}$ and (dense) $N_{\mathcal{H}} \times M$ matrices.
Notably, since for a linear parametrization $\dot{\S} = 0$ (see App.~\ref{App:EOM_derivation}), the computation of $\S^{-1}$ incurs no additional computational cost after its initial calculation at the beginning of the process. 

As previously anticipated, within each integration step, if $M\ll N_{\mathcal{H}}$ (LR hypothesis), both the speed and storage requirements for the calculation are dramatically improved. For additional information on the computational costs and procedures, we refer readers to App.~\ref{app:pseudoinverse}.

\subsection{Dynamical rank update}
\label{sec:dynamical_rank}
A critical aspect of any LR approach is its capability to dynamically adjust the dimension of the LR basis throughout the system's evolution [$M=M(t)$]. This adaptability is essential for accommodating changes in the system's entropy over time. 
In the ensemble truncation methods \cite{ciuti2021, chen2021, mccaul2021}, this update is achieved by truncating the full-rank density matrix $\rhoo(t+\dd t)$ at each time step with the goal of keeping a control quantity, namely the truncation error $\epsilon_M = 1-\sum_{j=1}^M p_j$, below a predefined threshold $\epsilon_{\rm max}$.
In this section we detail schemes for efficiently implementing a dynamic update of the dimension of the variational basis in our approach.

\subsubsection{Basis inflation}
Let's first consider the basis inflation problem, where the system's entropy increases over time. An example of a system presenting this behaviour is developed in Sec.~\ref{sec:XYZ}. To start, we posit the existence of a control quantity $\chi(t)$ that accurately reflects the solution's accuracy and that can be efficiently computed at each time step. Various options for $\chi$ will be discussed below. This quantity is monitored through a callback mechanism, continuously evaluated against a predefined upper bound of $\epsilon_{\rm max}$. Initially zero at $t=0$, $\chi$ grows with the system's entropy, and hence, over time.
Let $t^*$ denote the time when $\chi(t^*)=\epsilon_{\rm max}$. At this time, the rank is incrementally increased from $M$ to $M+1$ through the addition of a new state $\ket{\varphi_{M+1}}$ to the LR basis. The population and coherences associated with this new state are initially vanishing [$B_{j,M+1} = (B_{M+1,j})^{*}=0\,$], to ensure continuity in the solution. 
The variational principle ensures that the solution can be made independent of the specific choice of $\ket{\varphi_{M+1}}$. This independence, within the specified tolerance, is guaranteed if the state is added at a time prior to $t^*$, ensuring that $\ket{\varphi_{M+1}}(t^*)$ satisfies the TDVP at that time. 
To achieve this, we implement periodic checkpoints at intervals of $\Delta t$. Each time a threshold crossing is registered, the integration is restarted with an enlarged basis from the nearest checkpoint before the crossing. This approach also ensures that the bound on $\chi$ remains strict, as the response in $\chi$ may not be immediate following the basis inflation.
\begin{figure}[tb]
\center
% \hspace*{-1.em}
\includegraphics{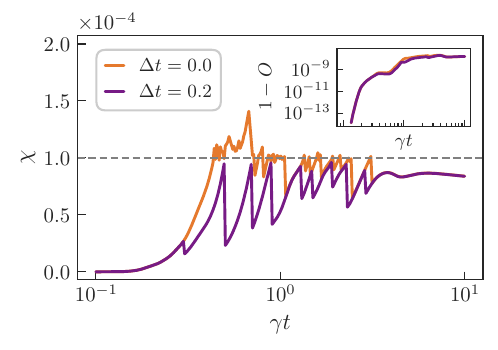}
\caption{\label{fig:chi_v_t}
Time evolution of the approximate variational error, $\chi(t)$, as defined in Eq.~\eqref{eqn:SiL}. The upper bound for $\chi$ is set at $\epsilon_{\rm max}=10^{-4}$, which strictly applies for $\Delta t>0$. The inset displays the overlap between the true solution and the truncated one as obtained from Eq.~\eqref{eqn:overlap}. The physical model underlying this example is the XYZ Heisenberg model (detailed in Sec.~\ref{sec:XYZ}) with $N=9$ spins and $J_y=1$. The values of all remaining physical parameters [see Eqs.~\eqref{eqn:XYZ_H}~and~\eqref{eqn:XYZ_L}] are the same as those used in Fig.~\ref{fig:xyz_magnetization}.}
\end{figure}
However, we observe that in most situations of interest, an instantaneous basis inflation ($\Delta t=0$) does not significantly compromise the solution's accuracy, suggesting that the variational adjustment of $\ket{\varphi_{M+1}}$ is almost instantaneous. This is illustrated in Fig.~\ref{fig:chi_v_t}, where we display $\chi(t)$ for both $\Delta t = 0$ and $\Delta t = 0.2$. As shown in the inset, the overlap between the exact and approximated solutions, evaluated using the expression
\begin{equation}
    \label{eqn:overlap}
    O(\hat A,\hat B) = \frac{\operatorname{Tr}\{\hat A^\dagger \hat B\}}{\sqrt{\operatorname{Tr}\{\hat A^\dagger\hat A\} \operatorname{Tr}\{\hat B^\dagger\hat B\}}},
\end{equation}
remains virtually unchanged when choosing $\Delta t= 0$.

\subsubsection{Basis deflation}
For the basis deflation problem, we adopt a similar approach to that used for basis inflation. Consider a case where the system's entropy, after peaking, gradually diminishes as the system approaches the stationary state. An example of such dynamics is discussed in Sec.~\ref{sec:TFIM}. In this context, to enhance efficiency, it is desirable to reduce the dimension of $\H_{M(t)}$ when $\chi$ falls below a lower threshold $\epsilon_{\rm min}$. Upon crossing this threshold, the rank is incrementally decreased from $M$ to $M-1$ by removing a state from the LR basis.
The selection of which state to remove from the basis is a critical aspect for the success of the deflation protocol. Intuitively, we should eliminate the state contributing the least to the dynamics, potentially the one with the lowest occupancy. However, a low occupancy does not necessarily imply negligible correlations or coherences. Therefore, we first transition to the diagonal basis by diagonalizing  $\rhot$ as 
\begin{equation}
\label{eqn:rho_diag}
\hat{\rho}(t)=\sum_{j=1}^Mp_j\ketbra{\eta_j}
\end{equation}
with $p_1\ge p_2\ge \ldots \ge p_M$. Subsequently, we remove the $M$th state from the basis but maintain the equivalent diagonal representation of the LR basis, resulting in $\z(t+\dd t) = (\ket{\eta_1},\ldots,\ket{\eta_{M-1}})$ and $\B(t+\dd t) = \operatorname{diag}(p_1,\ldots,p_{M-1})$.
Efficient diagonalization of $\rhot$ is feasible by recognizing that $\rhot$ can be expressed as $\hat{\rho}=\hat{C}\hat{C}^\dagger$, where $\hat{C}=\z\sqrt{\B}$. As discussed in Refs.~\cite{mccaul2021, ciuti2021, chen2021}, this matrix shares the same non-vanishing eigenvalues $p_j$ as the small $M\times M$ matrix $\hat{\sigma}=\hat{C}^\dagger\hat{C}$. The eigenvectors of these two matrices are related through $\hat C$.
It is important to note that, unlike the basis inflation case, failing to reduce the rank in the deflation scenario does not lead to a less accurate solution. Quite the contrary, the solution will be more accurate, albeit at a higher computational cost.

\subsubsection{Control quantities}
We now provide a set of possible choices for $\chi$. Given the variational parametrization of the LR states, the natural control quantity for our method is the distance between the true evolution and the variational evolution of the trial state, calculated as  \cite{benjamin2019}
\begin{equation}
    \chi = ||\dot\rhoo - \LL(\rhoo)||.
\end{equation} 
The only linear algebra operations involved in the computation of this quantity are matrix multiplications, the largest of which between extremely sparse $N_{\mathcal{H}} \times N_{\mathcal{H}}$ matrices and dense $N_{\mathcal{H}} \times M$ matrices. The number of matrix multiplications to be computed scales as $D^2$, the square of the number of jump operators.

A more efficient choice, which comes at no additional computational cost, is 
\begin{equation}
\label{eqn:SiL}
    \chi = \Tr{\S^{-1}\L} = \Tr{\P\LL(\rhoo)}.
\end{equation}
We adopted this choice for all simulations in the paper.
Although $\Tr{\LL(\rhoo)}=0$ for any physical density matrix $\rhoo$ and Liouvillian $\LL$, $\Tr{\P\LL(\rhoo)}$ can be non-zero as a result of projecting over an incomplete basis \cite{doriol2015, raab2000}.
Indeed, $\Tr{\P\LL(\rhoo)}$ is only vanishing when $\rhoo$ is entirely contained in the LR manifold. Assuming the latter to be constructed around the initial state of the simulation, $\Tr{\P\LL(\rhoo)}$ vanishes at $t=0$, and in an ideal evolution would remain so throughout the dynamics. Its departure from zero quantifies leakage outside of the variational manifold, and as such, it is a good indicator of the solution's accuracy in real time. 
This choice for $\chi$ is closely related to the truncation error $\epsilon_M$ of Ref.~\cite{ciuti2021}, as can be seen from the perturbative expansion carried out in App.~\ref{sec:DynCornerSpace}.

The last criterion we propose takes $\chi = p_M/p_1$ where the probabilities $p_i$ follow from the diagonalization of $\rhot$ according to Eq.~\eqref{eqn:rho_diag}.

\begin{figure}[tb]
\center
% \hspace*{-1.em}
\includegraphics[width=0.49\textwidth]{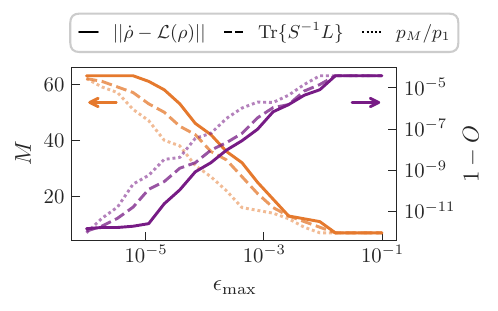}
\caption{\label{fig:rankscaling} 
Variation of the rank $M$ and of the overlap $O$ (between the true and approximated steady-state solutions) with the threshold $\epsilon_{\rm max}$. We present results for all the choices of control parameter $\chi$ discussed in the text. The underlying physical model is the XYZ Heisenberg model with $N=6$ spins and $J_y=1$. The remaining physical parameters take the same values as in Fig.~\ref{fig:xyz_magnetization}.}
\end{figure}

All three choices of $\chi$ are well suited control quantities as evidenced by Fig.~\ref{fig:rankscaling} where we display the overlap between the real and variational solution as a function of the threshold $\epsilon_{\rm max}$. As expected, $1-O$ [calculated from Eq.~\eqref{eqn:overlap}] and $M$ increase the lower the value of $\epsilon_{\rm max}$.

\subsection{Comparison with ensemble truncation methods}
\label{sec:comparison}
In this section, we outline the significant computational advantages that the LR-TDVP method offers over the ensemble truncation schemes discussed in Refs.~\cite{mccaul2021,chen2021,ciuti2021}:
\begin{enumerate}
    \item 
     Ensemble truncation methods execute linear algebra operations on $N_{\mathcal{H}} \times M(D+1)$ matrices, leading to larger computational and storage demands than those required by our variational LR density matrix approach, which only requires storing matrices of size at most $N_{\mathcal{H}} \times M$.
    \item Ensemble truncation methods rely on a Kraus representation of the Liouvillian map. The operators decomposing this map are proportional to $\sqrt{\dd t}$, thereby restricting these methods to the explicit Euler first-order integration scheme. In Ref.~\cite{ciuti2021}, this difficulty is partially lifted by replacing the Kraus operator $\hat{K}_0$ with the full time-evolution operator $\hat{K}_0=\operatorname{exp}(-iH_{\rm eff}\dd t)$ and using a high-order integration scheme for its application. The contribution of the other Kraus operators to the dynamics is however still limited to an explicit first-order in time Euler scheme. Conversely, our method allows integrating the EOM with a high-order adaptive-timestep solver. This approach not only significantly enhances the efficiency of our method but also offers greater accuracy and flexibility compared to the aforementioned schemes.
    \item Checks for basis inflation and deflation can be carried out at no additional costs to the simulation of our dynamics so that the additional diagonalization of an $M\times M$ matrix is only necessary when the deflation criterion is triggered. This is not the case in ensemble truncation methods which, for this purpose, require the diagonalization of an $M(D+1) \times M(D+1)$ matrix at every time step.
    \item Unlike ensemble truncation methods, where the selection of states to retain in the dynamics is heuristic in nature, our method's selection process is grounded in the time-dependent variational principle ensuring the adoption of the optimal LR basis at each time step.
\end{enumerate}

\section{Numerical simulations}
In this section, we outline the significant computational advantages that our variational method offers over the ensemble truncation schemes discussed in Refs.~\cite{mccaul2021,chen2021,ciuti2021}, and demonstrate its applicability in a broad range of physical models.

Although a key strength of our method lies in its ability to capture the entire Liouvillian dynamics, we begin our analysis focusing on systems for which the relevant physics occurs in the steady state. Here, we leverage our method only to reach stationarity, disregarding the remainder of the dynamics. In doing so, we showcase the method's effectiveness in analyzing steady-state phenomena in low-entropy systems.
In Sec.~\ref{sec:XYZ}, we investigate the dissipative phase transition in the anisotropic XYZ Heisenberg model. In Sec.~\ref{sec:FAF}, we apply our method to bosonic models, specifically to the study of the steady-state of the bosonic simulator of the triangular antiferromagnetic Ising model. This exploration underscores the versatility and general-purposefulness of our approach.

We proceed to showcase the algorithm’s capability to accurately capture the full Liouvillian dynamics, particularly in scenarios with non-monotonous entropy profiles.
In Sec.~\ref{sec:TFIM} we provide insight into the mechanisms of basis inflation and deflation by studying the dynamics of the transverse field Ising model in the presence of weak magnetic fields for which the system is known to be low-rank.
In Sec.~\ref{sec:cat_qubits}, we address the timely problem of simulating the full dynamics of bias-preserving gates in dissipative cat qubit architectures displaying once again the advantage of a LR approximation.

\subsection{Dissipative Heisenberg XYZ model}
\label{sec:XYZ}

We consider a two-dimensional lattice consisting of $N$ spin-1/2 particles governed by the Heisenberg XYZ Hamiltonian:

\begin{equation}
\label{eqn:XYZ_H}
\H = \sum_{\langle i,j\rangle} J_x \sx_i\sx_j + J_y \sy_i\sy_j + J_z \sz_i\sz_j + h_z\sum_j \sz_j,
\end{equation}
where $\hat\sigma_i^\alpha$ ($\alpha=x,y,z$) are the Pauli operators acting on the $i$th spin of the lattice. The incoherent relaxation of each spin is described by the dissipator of $\sm_j =(\sx_j-i\sy_j)/2$, leading to the master equation:
\begin{equation}
\label{eqn:XYZ_L}
\frac{\partial \rhoo}{\partial t} = \LL\rhoo = -i\comm*{\H}{\rhoo} + \gamma \sum_{j=1}^{N} \D[\sm_j]\rhoo.
\end{equation}

This equation exhibits a $\mathcal Z_2$ symmetry, as evidenced by its invariance under the transformation $\hat \sigma ^{x,y}_j\to-\hat \sigma ^{x,y}_j$ $\forall\,j$. In the thermodynamic limit, the steady state of the system bearing this symmetry is the fully aligned spin-down state, $\ket{\downarrow, \downarrow, \ldots, \downarrow}$. This state is characterized by having zero magnetization in the $xy$ plane, a defining feature of the so called \emph{paramagnetic phase}.

In the presence of anisotropic coupling in the $xy$-plane ($J_x\neq J_y$), and in the absence of any external field, however, relaxation and Hamiltonian processes compete with one another. This competition has been shown to induce a dissipative phase transition \cite{minganti2023, minganti2021, minganti2021a, minganti2023a, minganti2016, minganti2018, minganti2018a, minganti2021, bartolo2016, carmichael2016, kessler2013, lukin2013, benary2022, biella2016, sieberer2013, savona2017} spontaneously breaking the $\mathcal Z_2$ symmetry of the steady state and giving rise, in the thermodynamic limit, to an ordered phase with a non-vanishing in-plane magnetization, the so called \emph{ferromagnetic phase} \cite{biella2016, Rota2017, lukin2013, wouters2018, rossini2021}.

\begin{figure}[tb]
\center
% \hspace*{-1.5em}
\includegraphics{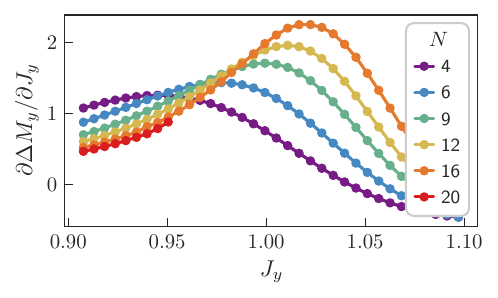}
\caption{\label{fig:xyz_magnetization} 
Derivative in $J_y$ of the variance $\Delta M_y$ of the magnetization along the $y$ axis for systems with different numbers of spins: $N=4$ ($2\times 2$ lattice), $N=6$ ($3\times 2$), $N=9$ ($3\times 3$), $N=12$ ($4\times 3$), $N=16$ ($4\times 4$), and $N=20$ ($5\times4$). 
Parameters: $J_x=0.9$, $J_z=\gamma=1$, and $h_x=0$. For $N<16$, where simulations are less demanding, we set $\epsilon_{\rm max}=10^{-5}$. For $N\geq16$, we set $\epsilon_{\rm max}=10^{-4}$.}
\end{figure}

In accordance with the conventions established in the relevant literature, all simulations of this model are performed fixing $J_x=0.9$, $J_z=\gamma=1$, and $h_z=0$. In this section we vary $J_y$ between $0.9$ and $1.1$ to investigate the transition from the paramagnetic to the ferromagnetic phase.
Observing the transition in a fully quantum model that retains all correlations in the system is challenging. Indeed, the steady-state magnetization
\begin{equation}
M_y = \expval*{\hat M_y}_{\rm ss} = \frac{1}{N}\sum_{j}\expval{\sy_j}_{\rm ss}
\end{equation}
along the $y$ direction, used as an order parameter in cluster mean-field studies \cite{biella2016}, is always zero in the full model. Similarly, the homogeneous steady-state structure factor
\begin{equation}
S^{xx} = \frac{1}{N(N-1)}\sum_{i\neq j} \expval{\sx_i\sx_j}_{\rm ss}
\end{equation}
which, in a Gutzwiller approximation is zero in the paramagnetic phase and acquires a positive value only in the ferromagnetic one \cite{wouters2018}, is not a good order parameter when correlations between different sites are included. 
Alternatives have been proposed. Specifically, the angle-averaged magnetic susceptibility \cite{Rota2017} and the trace-distance susceptibility \cite{Li2022}. 
The latter relies on the distance between steady state matrices evaluated at infinitesimally close values of $J_y$ and is thus too sensitive to any form of noise to be of use in approximated solutions. The former, however, would be well suited for our approximation. Its computation is however quite cumbersome as it requires assessing the linear response of the system to small perturbations. Here, we thus prefer to present as an indicator of the transition the variation in $J_y$ of
\begin{equation}
\label{eqn:Delta_M}
    \Delta M_y = \sqrt{\expval*{\hat M_y^2}} = \sqrt{\frac{1}{N^2}\sum_{i,j} \expval{\sy_i\sy_j}}\,\,,
\end{equation}
that is, the square root of the steady state expectation value of the variance of $\hat M_y$. This is a good indicator of the long-range correlations developing in the array across the transition, and its derivative $\partial \Delta M_y/\partial J_y$ in $J_y$ is a clear signature of critical behaviour. We display this quantity in Fig.~\ref{fig:xyz_magnetization} for increasingly larger system sizes.
A finite-size analysis in the linear lattice length $L=\sqrt{N}$ of the position of the maximum variance derivative yields an estimate of $J_c = 1.06\pm0.03$, consistent with the results obtained in Ref.~\cite{Rota2017}. The maxima of the curves themselves follow a power-law growth in $L$ with critical exponent $\kappa = 0.92 \pm 0.08$.

\begin{figure}[tb]
\center
% \hspace*{-1.5em}
\includegraphics{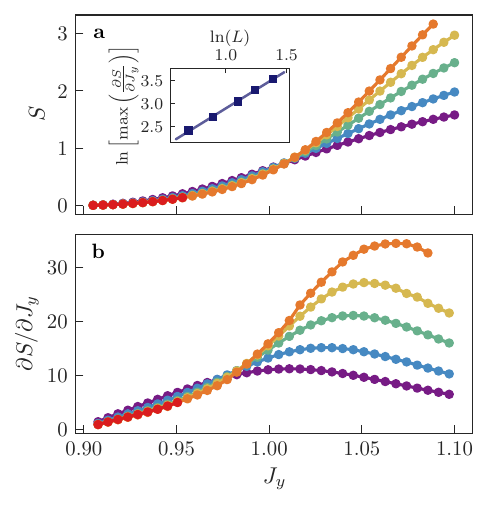}
\caption{\label{fig:xyz_entropy} 
Von Neumann entropy $S$ and its derivative $\partial{S}/\partial{J_y}$ as a function of the coupling parameter $J_y$ for lattices with different numbers of spins $N$ (legend in Fig.~\ref{fig:xyz_magnetization}). The inset displays the maximum value of the derivative as a function of the linear lattice length $L=\sqrt{N}$. The solid line is a power-law fit of the finite-size scaling.
We use the same physical parameters as in Fig.~\ref{fig:xyz_magnetization}.}
\end{figure}

To demonstrate the increasingly mixed character of the steady-state density matrix $\sss$, we display in Fig.~\ref{fig:xyz_entropy} the steady-state von Neumann
entropy
\begin{equation}
\label{eqn:entropy}
    S = -\Tr{\sss\ln(\sss)},
\end{equation}
and its derivative $\partial S/\partial J_y$ as a function of $J_y$. As expected, we find that in proximity of the critical point, the entropy sharply rises with a slope that increases with the system size. Concurrently, $\partial S/\partial J_y$ shows a peaked structure which becomes increasingly more pronounced as we move towards the thermodynamic limit. By fitting the maximum entropy derivative with a power law in $L$ [i.e $\max(\partial S/\partial J_y)\propto L^\lambda$], we get an estimate for the critical exponent of $\lambda = 1.6\pm0.1$, again consistent with the state-of-the-art results found in Ref.~\cite{Rota2017} via the corner space method.

The system's steady state is obtained by numerically integrating Eq.~\eqref{eqn:EOM} for a time $\gamma t=15$ long enough for the system to reach stationarity. In all simulations, the rank is dynamically increased starting from an initial choice of $M(0) = N+1$. As the initial state of the dynamics we take the pure state $\ket{\psi(0)} = \ket{\downarrow, \downarrow, \ldots, \downarrow}$ setting to zero the population of the remaining $N$ states making up the initial variational basis. These states are chosen as those with minimal Hamming distance from $\ket{\psi(0)}$.

\begin{figure}[tb]
\center
% \hspace*{-1.5em}
\includegraphics{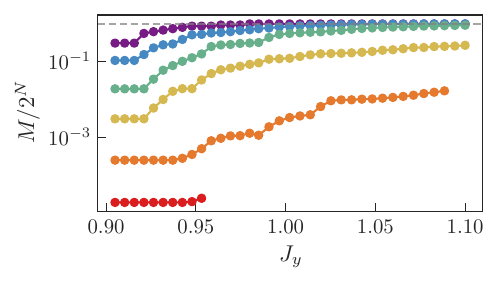}
\caption{\label{fig:xyz_rank}
Rank $M$ of the steady state of the system as a function of the coupling strength $J_y$. For small values of $J_y$, i.e small values of $S$, the rank of the system remains fixed at the initial preset value of $N+1$. The rank increases with the entropy of the system according to the discussion in Sec.~\ref{sec:dynamical_rank}.
We use the same physical parameters as in Fig.~\ref{fig:xyz_magnetization}.}
\end{figure}

As $S$ increases, additional states are required to accurately describe $\sss$, thereby progressively reducing the computational advantage introduced by a LR ansatz. Given that $S$ increases monotonically in $J_y$, the advantage introduced by our method is largest in proximity of the paramagnetic phase, where the entropy remains small. This is corroborated by Fig.~\ref{fig:xyz_rank}, where we display $M(\gamma t=15)$ as a function of $J_y$. Indeed, for highly entropic configurations [c.f.~Fig.~\ref{fig:xyz_entropy}(a)] $M\to2^{N}$, i.e the LR simulation becomes equivalent to the full solution of the master equation.

As a figure of merit for the method's computational performance we take the memory footprint of steady-state simulations similar to those performed above. The free parameters in this analysis are: 
\begin{itemize}
    \item The number of spins $N$, which uniquely determines the Hilbert space dimension.
    \item The coupling strength $J_y$, which uniquely determines the system's configuration. Recall that all other physical parameters have been fixed. 
    \item The dimension $M$ of the LR space. This can be fixed at the start, as we will do in what follows, or varied dynamically, in which case the free parameter becomes the threshold $\epsilon_{\rm max}$.
\end{itemize}

\begin{figure}[thb]
\center
% \hspace*{-1.5em}
\includegraphics[width=0.93\columnwidth]{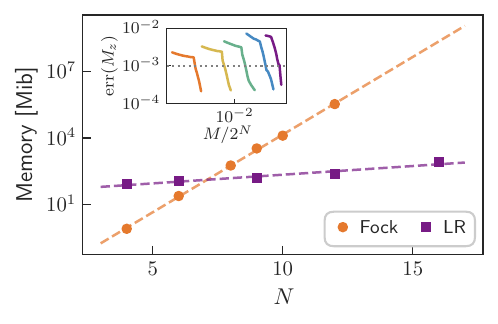}
\caption{\label{fig:xyz_mem} Memory allocations as a function of number of simulated spins $N$ for the full  simulation and the LR-TDVP one. We set the value of $J_y=0.98$. For each value of $N$, we choose the rank $M$ ensuring the relative error on the magnetization $M_z$ to be smaller than $0.1\%$, namely $\rm{err}(M_z) = |M_z^{\rm full} - M_z^{\rm LR}|/|M_z^{\rm full}| \leq 10^{-3}$. The inset shows the behaviour of $\rm{err}(M_z)$ as a function of $M$. The color scheme in the inset is the same as in Fig.~\ref{fig:xyz_magnetization}, as are all physical parameters aside from $J_y$.}
\end{figure}

For a synthetic two-dimensional representation of the results -- memory footprint against Hilbert space dimension -- as depicted in Fig.~\ref{fig:xyz_mem}, both $J_y$ and $M$ need to be fixed. 
We fix $J_y=0.98$ and select $M$, for each value of $N$, so as to ensure that the relative error in the estimation of $M_z$ remains within a $0.1\%$ margin from the actual solution. The distance between the LR estimate and the true value of $M_z$ for different lattice sizes is shown in the inset to Fig.~\ref{fig:xyz_mem}.
While for $N<16$ the true value of $M_z$ is obtained via numerical integration of the full master equation, for $N=16$ we resort to the Monte Carlo trajectory method, averaging over $10^4$ trajectories.
The results, as illustrated in Fig.~\ref{fig:xyz_mem}, highlight the significant computational advantages of our approach. The LR-TDVP method not only accurately captures all the critical features of the XYZ model but also demonstrates improved efficiency in terms of memory usage and computation time compared to both the full solution and the dynamical truncation schemes referenced in \cite{ciuti2021, mccaul2021, chen2021}.

In concluding this section, it is important to stress that our method is inherently dynamic in nature. As such, it is not expected to always outperform methods specifically tailored for solving steady-state problems. For instance, when compared to the corner space renormalization method, which serves as our reference, the LR-TDVP does not surpass it in terms of the system sizes that can be explored. The corner space method, based on spatial block decimation, is indeed very effective for systems with spatial translational invariance.

One of the key advantages of the LR-TDVP method over the corner space approach, however, is its generality and applicability to a broader range of low-entropy systems, not limited to those characterized by translational invariance. Additionally, this method offers more precise control over error estimation and convergence in the rank. In contrast, the corner space method's truncation at any decimation level introduces errors that, while locally controllable, can accumulate and propagate through different decimation steps, affecting the overall accuracy. This often necessitates running multiple simulations with varying truncation combinations to ensure convergence -- a requirement not present in our algorithm.

Furthermore, the merging process in the corner space method is not unique and can lead to artificially slow timescales, potentially affecting performance. The fixed but non-unique choice of basis in the corner space method depends on the initial lattice for decimation, meaning that missing a crucial state in the system can lead to erroneous or artificially slow simulations. Conversely, our algorithm allows for a more precise final result, even if part of the dynamics might be initially missed. Error control is much more straightforward, and can be dynamically adjusted by expanding the LR dimension. This flexibility offers a distinct advantage in terms of precision and control over the simulation process.

In summary, while the LR-TDVP method may not always surpass specialized steady-state solvers in terms of system size exploration, it compensates with its generality, error control, and adaptability. This makes it a valuable tool for studying a wide range of low-entropy quantum systems, including those with non-uniform interactions where methods like the corner space renormalization may not be applicable.

\subsection{Frustrated antiferromagnetism in quadratically driven QED cavities}
\label{sec:FAF}
We consider an array of $N=2,3$ coupled dissipative quadratically-driven QED nonlinear cavities: a bosonic model known to be a valid quantum simulator of the the triangular antiferromagnetic Ising model \cite{Wannier1950, rota2019a, rota2019, savona2017}. 
The Hamiltonian of the model reads  
\begin{equation}
    \begin{aligned}
        \hat{H}= & \sum_{j=1}^{N}-\Delta \hat{a}_j^{\dagger} \hat{a}_j+\frac{U}{2} \hat{a}_j^{\dagger 2} \hat{a}_j^2+\frac{G}{2} \hat{a}_j^{\dagger 2}+\frac{G^*}{2} \hat{a}_j^2 \\
        & -\sum_{j \neq j^{\prime}} \frac{J}{2}\left(\hat{a}_j^{\dagger} \hat{a}_{j^{\prime}}+\hat{a}_{j^{\prime}}^{\dagger} \hat{a}_j\right),
    \end{aligned}
\end{equation}
where $U$ is the Kerr nonlinearity, $G$ the two-photon driving field amplitude, $\Delta$ the resonators' detuning frequency, and $J$ the hopping coefficient. We include single- and two-body losses with rates $\gamma$ and $\eta$ respectively, so that the system evolves according to the Lindblad master equation
\begin{equation}
\label{eqn:Liouvillian_AF}
        \frac{\partial \rhoo}{\partial t} = \LL\rhoo = -i\comm*{\H}{\rhoo} + \sum_j \left(\gamma\D[\a_j] + \eta \D[\a_j^2] \right)\rhoo
\end{equation}

In the limit of a strong driving field, the photons in each cavity form a coherent state with phase $\alpha$ or $- \alpha$. While a positive hopping coefficient $J > 0$ favours the formation of a statistical mixture of two equiprobable separable coherent states \cite{gerry2004} $\ket{\Psi_{\pm}} = \prod_j \ket{\pm \alpha}_j$, a negative one $J < 0$ favours the emergence of states $\ket{\Phi_{\pm}} = \ket{\pm \alpha, \mp \alpha, \pm \alpha, \mp \alpha, \ldots}$ with an antisymmetric alignment of phases. The two limits map directly to the ferromagnetic and antiferromagnetic configuration of the effective spin model upon the mapping $\ket{\alpha}\to\ket{\uparrow}$ and $\ket{-\alpha}\to\ket{\downarrow}$.

\begin{figure}[tb]
\center
\hspace*{-1.5em}
\includegraphics{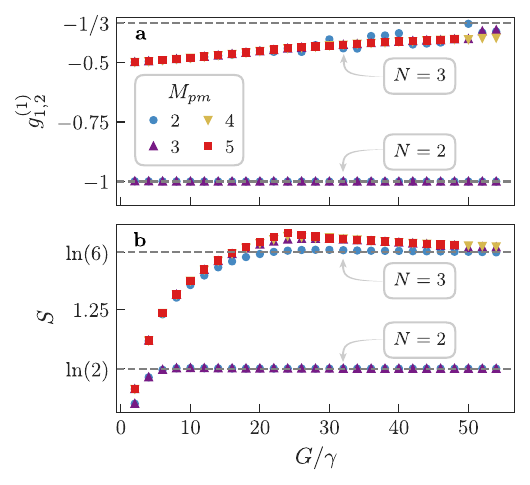}
\caption{\label{fig:FAF} 
The first-order correlation function $g_{1,2}^{(1)}$ (a) and the von Neumann entropy $S$ (b) as a function of the amplitude of the two-photon driving $G$ for the systems with $N = 2, 3$ cavities. Dashed grey lines are used to display the expected plateaux values of the different quantities in the limit of large driving.
Parameters: $\gamma=\eta=1$, $U=10$, $\Delta=J=-10$.}
\end{figure}

Following Ref.~\cite{rota2019a}, we investigate the effective antiferromagnetic spin model, by studying the steady-state properties of the photonic system for varying values of $G$. 
Specifically, we focus on the steady-state values of the first-order coherence correlation function
\begin{equation}
    g_{1,2}^{(1)}=\expval{\ad_1\a_2}\Big/\expval{\ad_1\a_1},
\end{equation}
and of the von Neumann entropy $S$ defined in Eq.~\eqref{eqn:entropy}. Our results, displayed in Fig.~\ref{fig:FAF}, are consistent with the findings in Ref.~\cite{rota2019a}. Indeed, as expected for an antiferromagnetic coupling, the correlation $g_{1,2}^{(1)}$ for $N=2$ is negative and, for increasing values of $G$, converges to the asymptotic value $g_{1,2}^{(1)}=-1$. Concurrently, the entropy increases from $S(G=0)=0$ to $S(G\gg \gamma) = \ln(2)$, consistently with the double degeneracy of the ground state manifold of the equivalent spin model. 

For an odd number of sites, the presence of geometrical frustration in the system hinders the emergence of $\ket{\Phi_\pm}$ states in the open system delaying to larger values of $G$ the onset of the antiferromagnetic signatures.
The effects of frustrations are visible for $N=3$ in the non-monotonous behavior of $S$ as a function of $G$, and in its asymptotic value of $S=\ln(6)$. Indeed, the latter is consistent with the existence of six possible spin configurations, emerging as a result of frustration, and minimizing the energy of the equivalent spin model.

We obtained all results by numerically integrating Eq.~\eqref{eqn:EOM} over a time interval sufficiently long to allow the system to reach the steady state. The Hilbert space of the $N$ cavities is obtained as the $N$-fold tensor product of truncated single-boson Fock spaces. For numerical convenience, we adjusted the truncation dimension according to the prescription $N_{cut}(G) = \max\{ 10\,,\,5 G/W\}$, where $W=\sqrt{U^2+\eta^{2}}$.
For $N=2$, a LR subspace of dimension $M=(M_{pm})^{N}$ with $M_{pm}=2$ (number of states per mode) accurately captures the relevant physics, as evidenced by the indistinguishable results obtained for $M_{pm}=3$. For $N=3$, a larger LR space is necessary. We present results for $M_{pm}=2,3,4,5$. Convergence is observed for $M_{pm}\geq3$.

Direct numerical integration of Eq.~\eqref{eqn:Liouvillian_AF} for determining the steady state of the system quickly becomes impractical, as highlighted by the extensive efforts in Ref.\cite{rota2019a}. In that approach, the steady state is computed by numerically solving $\partial_t \rhoo = 0$, a linear system of up to $10^{8}$ equations. In contrast, our method demonstrates significant potential by efficiently computing the steady state by integration of Eq.~\eqref{eqn:Liouvillian_AF} but retaining only a small number of optimal states. This approach not only reduces computational complexity but also maintains accuracy, showcasing the advantages of our method in handling large-scale quantum systems.

\subsection{Transverse field Ising model}
\label{sec:TFIM}
\begin{figure}[htb]
\center
\hspace*{-1.5em}
\includegraphics[width=0.52\textwidth]{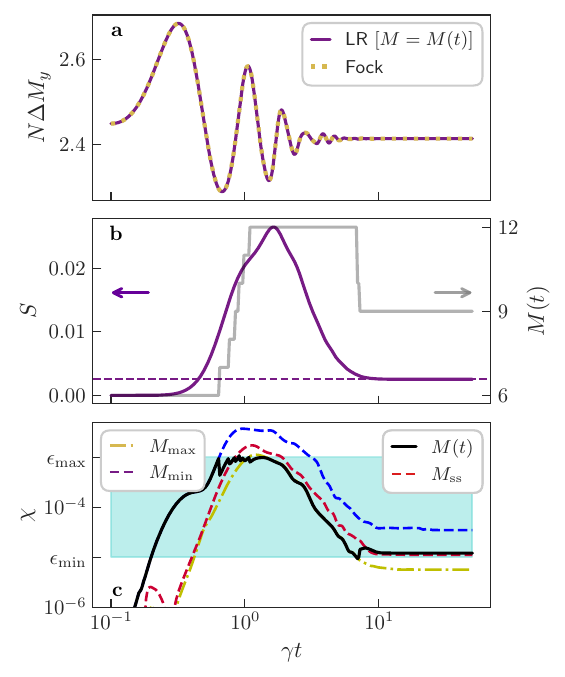}
\caption{\label{fig:Dynamics_example} 
Example of low-rank dynamics in the transverse field Ising model. (a) Time evolution of the standard deviation $\Delta M_y$ of the spin correlation along $y$. (b) Evolution of the system's entropy $S$ (left axis) and of the adaptive rank $M(t)$ (right axis). The dashed blue line indicates the true steady state value of $S$. (c) Evolution of the approximated variational error $\chi$ [Eq.~\eqref{eqn:SiL}] for different choices of the rank, both changing and static in time. The shaded blue region marks the interval $\epsilon_{\rm{min}} <\chi<\epsilon_{\rm{max}}$.
Parameters: $J_x=J_y=0$, $J_z=\gamma=1$, $h_x=0.75$, $N=6$, $\epsilon_{\rm min}=10^{-5}$, $\epsilon_{\rm max}=10^{-3}$, $M_{\rm min} = M(0) = 6$, $M_{\rm max} = 12 $, $M_{\rm ss} = M(\gamma t=50) = 9$.}
\end{figure}

In this section, we  demonstrate our algorithm's ability to accurately capture the full dynamics of a system while efficiently adapting to an entropy profile that does not monotonically increase over time. Specifically, we examine a scenario where the entropy initially rises, reaches a peak at $t_{\rm{max}}$ with a maximum value $S_{\rm{max}}$, and then decreases towards a steady-state value $S_{\rm{ss}}<S_{\rm{max}}$.

This behavior is exemplified in the transverse field Ising model [$J_x=J_y=0$, $J_z=\gamma=1$, and $h_x>0$ in Eqs.~\eqref{eqn:XYZ_H}~and~\eqref{eqn:XYZ_L}] which is characterized by a dynamics that, under specific initial conditions, remains LR at all times. Indeed, for sufficiently small values of $h_x$ in \eqref{eqn:XYZ_H}, the steady state of the model closely resembles that at $h_x=0$, which reads $\ket{\psi_0} = \ket{\downarrow}^{\otimes N}$. Consequently, an evolution under small $h_x$ starting from the latter state exhibits the desired entropy profile while remaining amenable to a LR representation. In contrast, an evolution starting from the state $\ket{\uparrow}^{\otimes N}$ would saturate the rank, rendering the LR ansatz an inefficient choice.

In Fig.~\ref{fig:Dynamics_example}(a), we display the time evolution of the structure factor $\Delta M_y$ as defined in Eq.~\eqref{eqn:Delta_M}. We remark the perfect agreement between the LR and full simulations.

Figure~\ref{fig:Dynamics_example}(b) depicts the desired evolution of the system's entropy and highlights the dynamic adaptation of the rank $M(t)$ to its changes throughout the simulation. 
Specifically, $M(t)$ initially increases, in parallel with $S$, to maintain the accuracy threshold set by $\epsilon_{\rm max}$. 
However, for $t > t_{\rm max}$, as $S$ decreases, $M$ is automatically reduced, thereby enhancing computational efficiency.

In Fig.~\ref{fig:Dynamics_example}(c), we display the variational error $\chi$, as defined in Eq.~\eqref{eqn:SiL}, which serves as a measure of the error performed at time $t$. As expected, the automatic adaptation of $M$ ensures that $\epsilon_{\rm{min}} <\chi(t)<\epsilon_{\rm{max}}$. 
In contrast, when the rank is fixed to either $M_{\rm max} = \max_{t}[M(t)]$ or $M_{\rm min} = \min_{t}[M(t)]$, there exist regions where $\chi(t)$ falls below $\epsilon_{\rm min}$ or exceeds $\epsilon_{\rm{max}}$, respectively.

\subsection{Cat qubit gates}
\label{sec:cat_qubits}
In this section we apply for the fist time a LR ansatz on dissipative cat qubit architectures \cite{gottesman2001, cochrane1999, goto2016, touzard2018, mirrahimi2014, guillaud2019, leghtas2015, albert2016, albert2018a, guillaud2022}, taking full advantage of the potentialities of our method to simulate bias-preserving logical operations \cite{albert2018a, mirrahimi2014, gilles1994, hachiii1994}. 

\subsubsection{Definition of dissipative cat qubits}
Schrödinger cat qubits \cite{mirrahimi2014, leghtas2015, albert2016, albert2018a, guillaud2022, puri2017, puri2019, puri2020, grimm2020, guillaud2019, venkatraman2022, chamberland2022, gautier2022, guillaud2021, frattini2022} encode information as a symmetric pattern in phase space, specifically, within the subspace $\{\ket{\C_{\alpha}^+}, \ket{\C_{\alpha}^-}\}$ spanned by cat states of opposite parity \cite{gilles1994, mirrahimi2014}, where 
\begin{equation}
\label{eqn:definition_cat}
\ket{\C_\alpha^\pm} = \frac{\ket{\alpha} \pm \ket{-\alpha}  }{2\sqrt{1 \pm e^{-2|\alpha|^2}}},
\end{equation}
and $\a \ket{\alpha} = \alpha \ket{\alpha}$ \cite{gerry2004, haroche2013}.
%Correspondingly, in the space of operators $\operatorname{Op}(\mathcal{H})$ acting on the bosonic Hilbert space $\mathcal{H}$, the analogous set is $\{\ketbra{\C_\alpha^+}{\C_\alpha^+}, \ketbra{\C_\alpha^+}{\C_\alpha^-}, \ketbra{\C_\alpha^-}{\C_\alpha^+}, \ketbra{\C_\alpha^-}{\C_\alpha^-}\}$. 
Within this manifold, the prevailing convention defines the logical states as 
\begin{equation}
    \begin{aligned}
        \ket{0} &= \frac{1}{\sqrt{2}}\qty(\ket{\C_\alpha^+} + \ket{\C_\alpha^-}) \approx \ket{+\alpha} + \order{e^{-2|\alpha|^2}}\\
        \ket{1} &= \frac{1}{\sqrt{2}}\qty(\ket{\C_\alpha^+} - \ket{\C_\alpha^-}) \approx \ket{-\alpha} + \order{e^{-2|\alpha|^2}},
    \end{aligned}
\end{equation}
so that $\ket{\pm} = (\ket{0} \pm \ket{1})/\sqrt{2} = \ket{\C_\alpha^\pm}$. In what follows we will assume $\alpha\in\mathbb{R}$.

Generating and stabilizing this manifold is far from trivial, as it requires engineering parity-preserving processes involving exclusively the pairwise exchange of photons between the system and its environment \cite{albert2018a, mirrahimi2014, gilles1994, hachiii1994}. 
In dissipative cat qubits, for instance, this manifold coincides with the four-degenerate steady-state manifold of the dissipator $\LL_0 = \kappa_2 \D[\a^2 - \alpha^2]$, modelling engineered two-photon drive and dissipation processes \cite{mirrahimi2014, leghtas2015, goto2016, cochrane1999, touzard2018, xu2022, gravina2023}. Indeed, because of the underlying strong $\mathcal{Z}_2$ symmetry of the Liouvillian, the right-kernel of $\LL_0$ is \cite{gravina2023, albert2014, albert2016, albert2018}
\begin{equation}
    \operatorname{As}(\mathcal{H}) = \lim_{t\to\infty} e^{\LL_0 t} \operatorname{Op}(\mathcal{H}) = \operatorname{Span}\left(\{\ketbra{\C_\alpha^\mu}{\C_\alpha^\nu}\}_{\mu,\nu=\pm}\right),
\end{equation}
so that, for any choice of $\rhoo_0$
\begin{equation}
    \sss = \lim_{t\to\infty}e^{\LL_0 t}\rhoo_0 = \sum_{\mu,\nu=\pm} c_{\mu,\nu} \ketbra{\C_\alpha^\mu}{\C_\alpha^\nu}.
\end{equation}
The coefficients $c_{\mu,\nu}$ encode the quantum information and are defined as 
\begin{equation}
    c_{\mu,\nu} = \Tr{\J_{\mu,\nu}^\dagger \rhoo_0},
\end{equation}
where the operators $\J_{\mu,\nu}$, spanning the left kernel of $\LL_0$, are defined as \cite{mirrahimi2014, albert2018}
\begin{equation}
\begin{aligned}
    \J_{++} &= \sum_{n=0}^\infty \ketbra{2n}{2n},\\
    \J_{--} &= \sum_{n=0}^\infty \ketbra{2n+1}{2n+1},\\
    \J_{+-} &= \J_{-+}^\dagger = \mathcal{A}_\alpha \sum_q a_q \J_{+-}^{(q)}.
\end{aligned}
\end{equation}
Here, 
\begin{equation}
\begin{aligned}
\mathcal{A}_\alpha &= \sqrt{\frac{2\alpha^2}{\sinh(2\alpha^2)}}, \qquad a_q = \frac{(-1)^q}{2q+1} I_q(\alpha^2),\\[0.3cm]
    \J_{+-}^{(q)} &= \left\{\begin{array}{lr}
    \dfrac{(\ad\a-1)!!}{(\ad\a+2q)!!} \, \J_{++} \, \a^{2q+1} & (q \geq 0)\\[0.3cm]
    \J_{++} \, (\ad)^{2|q|-1} \, \dfrac{(\ad\a)!!}{(\ad\a+2|q|-1)!!}\,\, & (q < 0)
    \end{array}\right.,
    \quad
\end{aligned}
\end{equation}
where $I_q(\cdot)$ is the modified Bessel function of the first kind, and $n!! = n \cdot (n-2)!!$ is the double factorial, applied element-wise to all diagonal operators.

The two-photon drive and dissipation processes modelled by $\LL_0$ have been realized, for example, on superconducting circuit platforms. In these platforms, the dominant source of errors comes from single photon loss processes, modeled by the dissipator $\kappa_1 \D[\a]$ with $\kappa_1/\kappa_2\ll 1$. These processes hinder the code's ability to encode quantum information as well as the performance of logical gates \cite{joshi2021, xu2022, gautier2022}.

\subsubsection{$Z$, $ZZ$, and $ZZZ$ gates}
Let us consider the application of $Z$ rotations of an angle $\pi$ over $N=1,2,3$ cat qubits. For a single qubit, these rotations are equivalent to the application of a logical Pauli $\sz$ operator, which introduces a phase of $(-1)$ on the logical $\ket{1}$ state \cite{nielsen2011}. In the cat basis, this amounts to exchanging $\ket{\C_\alpha^+}$ and $\ket{\C_\alpha^-}$. 
Similarly, in multi-qubit systems, these rotations correspond to a unitary transformation changing the tensor product $\ket{\C_\alpha^\pm}\ldots\ket{\C_\alpha^\pm}$ into $\ket{\C_\alpha^\mp}\ldots\ket{\C_\alpha^\mp}$.

Logical $Z$ operations can be approximately implemented by evolving the system for a time $T$ under the Liouvillian $\LL = \LL_0 + \LL_1$ given by \cite{xu2022, gautier2022, mirrahimi2014, guillaud2021, guillaud2019}: 
\begin{equation}
    \begin{aligned}
        \LL_0 &= \sum_{i=1}^{N} \kappa_2 \D[\a_i^2 - \alpha_i^2],\\
        \LL_1 \rhoo &= -i\comm{\H_Z}{\rhoo} + \sum_{i=1}^{N} \kappa_1 \D[\a_i]\rhoo.
    \end{aligned}
\end{equation}
Here, $\LL_1$ encompasses both the unwanted single body losses that compromise the accuracy of gate operations, and the ideal Hamiltonian evolution responsible for the rotation. Specifically, the Hamiltonian $\H_Z$ is defined as \cite{chamberland2022, regent2023}:
\begin{equation} 
\H_Z = 
    \left\{\begin{array}{lr}
        \epsilon_Z    (\a_1 + \ad_1) & (N=1)\\
        \epsilon_{ZZ} (\a_1\ad_2 + \ad_1\a_2) & (N=2)\\
        \epsilon_{ZZZ} (\a_1\a_2\ad_3 + \ad_1\ad_2\a_3) \,\, & (N=3)
    \end{array}\right. 
    \quad
\end{equation}
with 
\begin{equation}
    \epsilon_{Z} = \frac{\pi}{4\alpha T}, \quad \epsilon_{ZZ} = \frac{\pi}{4\alpha^2 T}, \quad \epsilon_{ZZZ} = \frac{\pi}{4\alpha^3 T}\,\,.
\end{equation}

\begin{figure}[tb]
\center
\hspace*{-2.em}
\includegraphics{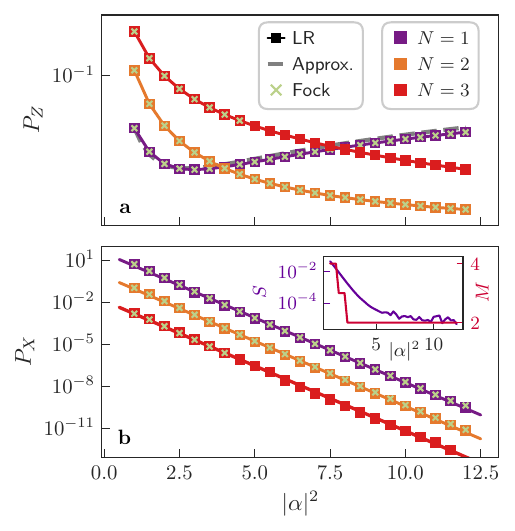}
\caption{\label{fig:Zgate} Error probabilities of $N$-qubit $Z$ gates, with $N=1,2,3$. (a) Phase-flip error probability $P_Z$ [Eq.~\eqref{eqn:PZ}] as a function of the photon number $\alpha^2$.
We display as a dashed gray line the analytical approximation $P_Z\approx \kappa_1T\alpha^2 + \epsilon_Z^2T/\alpha^2$ (for $N=1$) provided in Ref.~\cite{chamberland2022}. 
(b) Bit-flip error probability $P_X$ [Eq.~\eqref{eqn:PX}] as a function of the photon number $\alpha^2$. To improve visualization, as the curves would largely overlap, we multiply each curve by a different constant, effectively shifting them vertically on the plot. The constants are $R = 10^2, 1, 10^{-2}$ for $N=1,2,3$ respectively. Full lines are used to display the exponential fit of the LR data. Because of computational constraints, the exact simulations for $N=3$ were only possible up to $\alpha^2=5$. The inset shows the von Neumann entropy $S$ at time $T$ as a function of $\alpha^2$ for $N=1$.
In both panels, the results obtained from the L simulations (squares) are in agreement with the full solution (yellow crosses).
Parameters: $\kappa_2 = 1$, $\kappa_1=1/1000$, $\epsilon_{Z} = \epsilon_{ZZ} = \epsilon_{ZZZ} = 1/20$.}
\end{figure}

If we initialize each qubit in the $\ket{+}$ state, after a time $T$, ideally, all qubits should be in the $\ket{-}$ state, for which $\mel{-}{\J_{++}}{-}=0$. Therefore, we can use
\begin{equation}
\label{eqn:PZ}
    P_Z = \left.\expval{\bigotimes_{i=1}^{N}\J_{++}}\right|_{t=T}
\end{equation}
as a measure of the  phase-flip error probability during the $Z$ rotation. 
Fig.~\ref{fig:Zgate}(a) displays $P_Z$ as a function of $\alpha^2$. Simulations were conducted in a truncated Fock space with dimension $N_{cut}(\alpha) = \max\{20\,,\, \ceil{4.5 \alpha^2}\}$, ensuring the convergence of each data point. 
Notably, the L simulations, executed with $M_{pm}=3$ states per mode, overlay perfectly with the results obtained from the full evolution. 
Simulations for $N=3$ were restricted to $\alpha^2<5$ because of prohibitively large memory requirements.

Conversely, since any rotation leaves the states lying on the rotation axis unchanged, if we initialize each qubit in $\ket{0}\approx\ket{\alpha}$, we expect to find them in that exact same state after evolving under $\LL$ for a time $T$.
As a quantifier of the bit-flip error rate affecting the system we can therefore take
\begin{equation}
\label{eqn:PX}
    P_X = \left.1-\expval{\bigotimes_{i=1}^{N}\operatorname{sgn}(\hat x)}\right|_{t=T},
\end{equation}
where $\hat x = \a + \ad $, and $\operatorname{sgn}(\hat x) \approx \J_{+-}$.
In Fig.~\ref{fig:Zgate}(b) we display $P_X$ as a function of $\alpha^2$. Noticeably, our model is able to accurately capture $P_X$  even up to the very small error probabilities associated with its exponential suppression in $\alpha^2$ \cite{mirrahimi2014, leghtas2015, albert2016, albert2018a}. The suppression coefficients $\zeta$ extracted from an exponential fit of the exact simulations are $\zeta = 2.13\pm 0.01$, $2.13\pm 0.01$, and $ 2.16\pm 0.03$, respectively for $N=1,2,3$. The same fit on the LR data 
yields $\zeta = 2.13\pm 0.01$, $2.14\pm 0.01$, and $ 2.14\pm 0.02$. These estimates are in agreement with each other and with the theoretical expectation \cite{chamberland2022}. As we show in the inset of Fig.~\ref{fig:Zgate}(b), the entropy $S$ of the target state at time $T$ decreases with the number of photons in the system. This entails that progressively less states are necessary to capture the dynamics. To avoid over-representation in our LR ansatz we thus reduce $M_{pm}$ (number of states per mode included in the simulation) with the photon number. Specifically, we obtain good agreement with the full solution by taking $M_{pm}=1$ for $\alpha^2>6$. 

\section{Conclusions}
\label{Sec:Conclusions}
In this paper, we introduce a novel, model-independent method designed for the efficient simulation of the dynamics of low-entropy systems. Recognizing that such evolution can be accurately captured by a limited number of states, our method builds upon and advances the previously established ensemble truncation schemes \cite{ciuti2021, chen2021, mccaul2021, rouchon2013}, integrating their key features within the framework of the time-dependent variational principle recently developed for open quantum systems \cite{weimer2015, doriol2014, doriol2015, benjamin2019}. Our approach enhances previous ensemble truncation methods by offering a rigorous and systematic protocol for defining and dynamically modifying the low-rank basis. Furthermore, it extends prior variational descriptions of dissipative systems by introducing a computationally efficient protocol for adapting the size of the variational manifold, thereby dynamically adjusting the rank of the density matrix throughout the simulation. This dynamic adaptation ensures that the low-rank subspace of the Hilbert space optimally represents the system's state at all times.

We have conducted extensive benchmarking of our method across various model systems that have garnered significant attention in recent decades. Particularly noteworthy is its application to multi-qubit bosonic codes \cite{mirrahimi2014, leghtas2015, albert2016, albert2018a, guillaud2022, puri2017, puri2019, puri2020, grimm2020, guillaud2019, venkatraman2022, chamberland2022, gautier2022, guillaud2021, frattini2022}, which are emerging as promising candidates for fault-tolerant quantum hardware.

To ensure easy integration into diverse research workflows, we have implemented our method in Julia \cite{bezanson2017julia} and incorporated it into the QuPhys library \cite{quphys,git}, a comprehensive toolkit for quantum system simulation.

Looking ahead, potential extensions of our work include the incorporation of efficient variational representations of the low-rank space, such as tensor-network \cite{Orus_Tensor_2019,cirac2021} or neural quantum states \cite{Carleo_Neural_2017}. These advancements could enable the simulation of larger-scale systems. Additionally, a digital version of our method could approximate the simulation of noisy quantum computing hardware, potentially enhancing error-mitigation protocols that rely on estimating hardware-specific noisy quantum channels \cite{vandenBerg2023, Temme2017}.

\begin{acknowledgments}
We acknowledge several fruitful discussions with Fabrizio Minganti, Alberto Mercurio, and Lorenzo Fioroni. We are grateful to Tim Besard for providing useful insight into the GPU porting of the method.
This work was supported by the Swiss National Science Foundation through Projects No. 200020\_185015, 200020\_215172, and 20QU-1\_215928, and was conducted with the financial support of the EPFL Science Seed Fund 2021. 
\end{acknowledgments}

\appendix

\section{Derivation of Eq.~\eqref{eqn:EOM}}
\label{App:EOM_derivation}

Given the generic parametrization $\rhoo(t) = \rhoo(\btheta(t))$, the EOM for the parameters $\btheta$ that at each time best approximate the action of $\LL$ on $\rhoo$, are obtained by minimizing the Frobenius distance between the variational evolution and the physical one. This is ensured by requiring the variation of this quantity to vanish, that is, by imposing
\begin{equation}
\label{eqn:MacLachlan_no_constraints}
    \delta\,\norm{\left(\dv{t}-\LL\right)\rhoo\,}^2=0.
\end{equation}
Note that in this notation the time derivative $\dd/\dd t$ is intended with respect to the variational parameters, i.e 
\begin{equation}
\dot\rhoo = \frac{\dd \rhoo}{\dd t}  = \grad_{\btheta}\rhoo(\btheta)\cdot\bthetadot = \sum_j \frac{\partial \rhoo}{\partial \theta_j}\dot\theta_j.
\end{equation}
With this definition, Eq.~\eqref{eqn:MacLachlan_no_constraints} can be rewritten as (see Ref.~\cite{benjamin2019} for the full derivation)
\begin{equation}
\begin{aligned}
\label{eqn:MacLachlan_no_constraints_expanded}
    0 &= \delta \left[ \Tr{\qty(\dot\rhoo-\LL(\rhoo))^\dagger \qty(\dot\rhoo-\LL(\rhoo))} \right]\\
    &=\Tr{\left(\frac{\partial \rhoo}{\partial \theta_i}\right)^\dagger \left(
    \sum_j \frac{\partial \rhoo}{\partial \theta_j}\dot\theta_j - \LL(\rhoo)\right)}\delta\dot\theta_i^{\,*} + \rm{h.c},
\end{aligned}
\end{equation}
so that 
\begin{equation}
\label{eqn:MacLachlan_no_constraints_full}
    \Tr{\left(\frac{\partial \rhoo}{\partial \theta_i}\right)^\dagger \left(
    \dot\rhoo - \LL(\rhoo)\right)} = 0
\end{equation}
for all choices of $i$. This is equivalent to the Dirac and Frenkel variational principle, where the projection of the stochastic evolution onto the tangent subspace $\{\partial \hat\rho/\partial \theta_i\}_i$ is made to vanish. 

Equation~\eqref{eqn:McLachlan} results from incorporating the Lagrange multiplier $\lambda$ into Eq.~\eqref{eqn:MacLachlan_no_constraints}, thereby embedding trace preservation directly into the EOM. Analogous calculations as those presented above lead from Eq.~\eqref{eqn:McLachlan} to 
\begin{equation}
\label{eqn:MacLachlan_with_constraints}
    0 = \Tr{\left(\frac{\partial \rhoo}{\partial \theta_i}\right)^\dagger \left(
    \sum_j \frac{\partial \rhoo}{\partial \theta_j}\dot\theta_j - \LL(\rhoo) + \lambda\mathds{1}\right)}\delta\dot\theta_i^{\,*} + \rm{h.c},
\end{equation}
where choosing the constraint $\dv{t}\Tr{\rhoo + \rhoo^\dagger}$, over the more obvious choice $\dv{t}\Tr{\rhoo}$, preserves the symmetry under complex conjugation found in Eq.~\eqref{eqn:MacLachlan_no_constraints_expanded}. Because of this, Eq.~\eqref{eqn:MacLachlan_with_constraints} unambiguously results in
\begin{equation}
\label{eqn:MacLachlan_with_constraints_full}
    \Tr{\left(\frac{\partial \rhoo}{\partial \theta_i}\right)^\dagger \left(
    \dot\rhoo - \LL(\rhoo) + \lambda\mathds{1}\right)} = 0
\end{equation}
for all choices of $i$.

Consider now the specific LR ansatz in Eq.~\eqref{eqn:LRansatz}, for which the variational parameters are the populations $B_{ij}$ and the coefficients $z_{\alpha k}$. For the populations $B_{ij}$, Eq.~\eqref{eqn:MacLachlan_with_constraints_full} amounts to
\begin{equation}
\label{eqn:EOMB_extended}
    \begin{aligned}
        0 &= \Tr{\left(\frac{\partial \rhoo}{\partial B_{ij}}\right)^\dagger \left(
    \dot\rhoo - \LL(\rhoo) + \lambda\mathds{1}\right)}\\[0.1cm]
    &=\mel{\varphi_i}{\dot\rhoo - \LL(\rhoo) + \lambda\mathds{1}}{\varphi_j} \\
    &= \mel{\varphi_i}{\dot\rhoo - \LL(\rhoo)}{\varphi_j} + \lambda \S_{ij}\\
    &= \qty(\S\Bdot\S + \btau\B\S + \S\B\btau^\dagger -\L)_{ij} + \lambda \S_{ij},
    \end{aligned}
\end{equation}
where the last equality follows from substituting
\begin{equation}
\label{eqn:rho_dot}
    \dot\rhoo = \sum_{l,m} \dot{B_{lm}}\ketbra{\varphi_l}{\varphi_m} + B_{lm}\ketbra{\dot \varphi_l}{\varphi_m} + B_{lm}\ketbra{\varphi_l}{\dot \varphi_m},
\end{equation}
as detailed in the calculations in Ref.~\cite{doriol2015}.
Upon isolating $\Bdot$ in Eq.~\eqref{eqn:EOMB_extended}, we get
\begin{equation}
\label{eqn:Bdot_full}
    \Bdot = \Si \L \Si - \left(\Si\btau\B + \B\btau^\dagger\Si \right) -\lambda \Si,
\end{equation}
where $\btau = \z^\dagger \zdot$ and, as explained in the main text, $\Si$ and $\Bi$ are the pseudoinverse matrices of respectively $\S$ and $\B$.
The value of $\lambda$ is determined by the constraint
\begin{equation}
    \begin{aligned}
        0 &= \dv{t}\Tr\{\rhoo\} = \Tr{\Bdot\S + (\btau+\btau^\dagger)\B} \\
        &= \Tr{\S^{-1}\L - \lambda},
    \end{aligned}
\end{equation}
from which $\lambda = \Tr{\S^{-1}\L}/\Tr{\mathds{1}}=\Tr{\S^{-1}\L}/M$. 

Equation~\eqref{eqn:Bdot_full} is general, and holds for any analytical parametrization of the LR states. The variational equation for the coefficients $z_{\alpha k}$ as presented in Eq.~\eqref{eqn:MacLachlan_with_constraints_full}, on the other hand, is explicitly dependent on the choice of parametrization. For the full parametrization in Eq.~\eqref{eqn:phi_LinearAnsatz}, in particular, it can be similarly reformulated as
\begin{equation} 
\label{eqn:EOMz_extended}
    \begin{aligned}
        0 &= \Tr{\left(\frac{\partial \rhoo}{\partial z_{\alpha k}^*}\right)^\dagger \left(
    \dot\rhoo - \LL(\rhoo) + \lambda\mathds{1}\right)}\\[0.1cm]
    &=\sum_{j,\beta}\mel{e_\alpha}{\dot\rhoo - \LL(\rhoo) + \lambda\mathds{1}}{e_\beta} B_{jk}z_{\beta j}\\
    &=\sum_{j,\beta} B_{jk}z_{\alpha j} \mel{e_\alpha}{\dot\rhoo - \LL(\rhoo)}{e_\beta} + \lambda(\z\B)_{\alpha k}\\
    &= \qty[ \z\Bdot\S\B + \zdot\B\S\B + \z\B\zdot^\dagger\z\B -(\L-\lambda \mathds{1})\z\B ]_{\alpha k},
    \end{aligned}
\end{equation}
where the last equality follows once again from Eq.~\eqref{eqn:rho_dot}. Upon multiplying by $\B^{-1}$ on the right, and substituting Eq.~\eqref{eqn:Bdot_full} for $\Bdot$, we find that
\begin{equation}
    (\mathds{1} - \P)\zdot\B\S = \qty( \Ltil -\z\S^{-1}\L ),
\end{equation}
where $\P = \z\S^{-1}\z^\dagger$ is the projector on the LR manifold. Since by definition $\P\zdot=0$, the latter reduces to
\begin{equation}
\label{eqn:zdot}
    \zdot = \qty(\Ltil- \z\Si\L)\Si\Bi.
\end{equation}

A few comments on the full parametrization are now in order. First, it is evident from Eq.~\eqref{eqn:zdot} that $\btau = \z^\dagger\zdot = 0$, which consequently reduces Eq.~\eqref{eqn:Bdot_full} to the form presented in Eq.~\eqref{eqn:EOM}. 
Second, although Ref.~\cite{doriol2015} showed that using a Lagrange multiplier method to enforce both energy and trace preservation does not, in general, result in the unitary evolution expected for an isolated system; any linear parametrization as the one in Eq.~\eqref{eqn:phi_LinearAnsatz} is exempt from this problem.

\section{Computational details}
\label{app:pseudoinverse}

The integration of the evolution problem presented in Eq.~\eqref{eqn:EOM} relies exclusively on linear algebra operations involving matrices of dimensions $N_{\mathcal{H}} \times M$ ($\z$ and $\Ltil$) and/or $M\times M$ ($\S$, $\L$, and $\B$), where $M\ll N$. This disparity in dimensions allows for efficient execution of these operations, given the relatively smaller size of at least one dimension of each matrix. The bulk of the computational effort is dedicated to calculating the matrix $\Ltil$, defined as:

\begin{equation}
\begin{aligned}
    \Ltil &= \LL(\rho)\z = \left[-i\left(\H_{\rm eff} \rhoo -\rhoo \H_{\rm eff}^\dagger\right) + \sum_{\sigma=1}^D \hat \Gamma_\sigma \rhoo \,\Gamma_\sigma^\dagger\right]\z\\[0.15cm]
    & = -i\qty(\H_{\rm eff} \z\B)\S +i \z\qty[\qty(\H_{\rm eff} \z \B)^\dagger\z] \,\,+ \\
    &\quad + \sum_{\sigma=1}^D \left(\hat \Gamma_\sigma \z\right)\B\qty[\left(\hat \Gamma_\sigma \z\right)^\dagger\z].
    \end{aligned}
\end{equation}

By following the order of operations as indicated by the parentheses in the equation, the computation of $\Ltil$ involves matrix-matrix multiplications between $M\times M$ matrices, $N_{\mathcal{H}} \times M$ and $M\times M$ matrices, as well as between extremely sparse $N_{\mathcal{H}} \times N_{\mathcal{H}}$ and dense $N_{\mathcal{H}} \times M$ matrices. This approach ensures that no dense $N_{\mathcal{H}} \times N_{\mathcal{H}}$ matrix is ever stored in memory or involved in any multiplication. The process is optimized for efficiency, leveraging the reduced dimensionality to minimize the computational load, and free from the diagonalization of large matrices required in Refs.~\cite{ciuti2021, mccaul2021, chen2021}.

\begin{figure}[htb]
\center
% \hspace*{-1.em}
\includegraphics{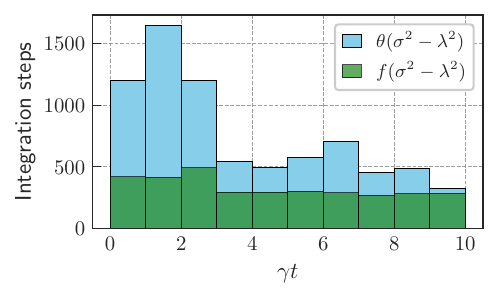}
\caption{\label{fig:pinv_regularization} Number of integration steps in of the adaptive integrator as a function of time. We simulate the evolution of the XYZ model with $N=9$ spins and $J_y=1$. The remaining physical parameters are set to the values discussed in Sec.~\ref{sec:XYZ}. We evolve while dynamically adjusting the rank with a threshold of $\epsilon_{\rm max}=10^{-4}$. The regularization parameters are: $\text{atol}=10^{-6}$ and $\text{rtol}=10^{-5}$}
\end{figure}

The integration of Eq.~\eqref{eqn:EOM} does however necessitate the calculation of the inverse matrices of $\S$ and $\B$. Since these matrices may be singular, their inverse is often ill-defined. In order to progress, regularization schemes are typically employed. A common approach involves adding a small diagonal contribution to the matrices to mitigate the impact of small eigenvalues. 
Yet, we have found that the effect of such an addition is not well controlled during time evolution, particularly when using an adaptive integrator. Similar observations were reported in Ref.~\cite{medvidovic2023}. Taking inspiration from this work, we adopt a regularization scheme based on the singular value decomposition of $\S = \bm U \bm \Sigma \bm V^\dagger$, where $\bm \Sigma = \operatorname{diag}(\sigma_1^2,\ldots,\sigma_M^2)$. Specifically, we define the pseudo-inverse of $\S$ as 
$\S^{-1} = \bm V \bm \Sigma^+ \bm U^\dagger$ where 
\begin{equation}
\label{eqn:pinv}
\Sigma_{\mu \nu}^{+}=\frac{f\qty(\sigma_\mu^2)}{\sigma_\mu^2} \, \delta_{\mu \nu}\quad\text{with}\quad f\qty(\sigma^2)=\frac{1}{1+\left(\frac{\lambda^2}{\sigma^2}\right)^6}.
\end{equation}
We do the same for $\B$. The value of $\lambda$ in chosen adaptively at each iteration according to 
\begin{equation}
\lambda^2=\lambda^2\left(\sigma_1^2, \ldots, \sigma_M^2\right)=\max \left(\text {atol} \,,\, \operatorname{rtol} \times \max _\mu\left(\sigma_\mu^2\right)\right).
\end{equation}
Throughout the paper we set $\text{atol}=10^{-6}$. 
Note that the Moore-Penrose pseudo-inverse corresponds to choosing the discontinuous function $f(\sigma^2) = \theta(\sigma^2-\lambda^2)$.  In line with the results from Ref.~\cite{medvidovic2023}, we find that opting for a smooth functional form for $f(\sigma^2)$ significantly enhances the stability and efficiency of the adaptive time-stepping in the integration routine. This improvement is illustrated in Fig.~\ref{fig:pinv_regularization}, where we plot the number of integration steps required as a function of time.

\section{The low-rank dynamical truncation scheme}
\label{sec:DynCornerSpace}
We review here the ensemble truncation methods (or \emph{low-rank dynamical truncation schemes}) introduced in Refs.~\cite{mccaul2021,chen2021,ciuti2021}. The starting assumption is that, at time $t$, the system is described by a LR density matrix in diagonal form
\begin{equation}
    \label{eqn:diagrhot}
    \hat{\rho}(t)=\sum_{j=1}^{M(t)}p_{j}\ketbra{\varphi_j}\,,
\end{equation}
where $\braket{\varphi_j}{\varphi_k}=\delta_{jk}$, and $M(t)\equiv M$ for notational clarity. The evolved state $\hat{\rho}(t+dt)$ can be written in terms of the Kraus operators as
\begin{equation}
    \label{eqn:rhotdt}
    \hat{\rho}(t+\dd t)=\sum_{j=0}^D\hat{K}_j\hat{\rho}(t)\hat{K}_j^\dagger+\order{\dd t^2}\,,
\end{equation}
where 
\begin{equation}
\begin{aligned}
\hat{K}_0=\I-i\H_{\rm eff}\dd t,\,\,\,  
&& \H_{\rm eff} = \hat{H} - \frac{i} {2}\sum_{\sigma=1}^D\hat{\Gamma}_\sigma^\dagger\hat{\Gamma}_\sigma,
\end{aligned}    
\end{equation}
and 
\begin{equation}
\hat{K}_\sigma=\hat{\Gamma}_\sigma\sqrt{\dd t}\quad \text{for} \quad \sigma=1,\ldots,D.  
\end{equation} 

Both $\hat{\rho}(t)=\bm{C}\bm{C}^\dagger$ and $\rhoo(t+\dd t)=\bm{T}\bm{T}^\dagger$ are then conveniently rewritten in terms of
\begin{equation}
    \begin{aligned}
        &\bm C = [\sqrt{p_j}z_{\alpha j}] && \text{with} && \dim(\bm C) = N_{\mathcal{H}} \times M,\\
        &\bm T = [\mel{e_\alpha}{\hat{K}_\sigma}{\varphi_k}\,] && \text{with} && \dim(\bm T) = N_{\mathcal{H}} \times M(D+1),
    \end{aligned}
\end{equation}
where $\sigma=0,\ldots,D$, $k=1,\ldots,M$, and $l=M\sigma+k$.

The dynamical truncation scheme consists, at each time step, in (i) computing the matrix $\bm{T}$, (ii) diagonalizing $\hat{\rho}(t+\dd t)$, and (iii) truncating the subspace by retaining only the eigenstates with the largest eigenvalues $\tilde{p}_j$. This last step determines the new rank $M(t+\dd t)$, according to the criterion 
\begin{equation}
\label{eqn:truncation_error}
    \epsilon_{M(t+\dd t)} \equiv \qty(1-\sum_{j=1}^{M(t+\dd t)}\tilde{p}_j)\le\epsilon_{\rm max}\ll1,
\end{equation}
where $\epsilon_{\rm max}$ is the upper bound discussed in Sec.~\ref{sec:dynamical_rank}. The new truncated density matrix reads
\begin{equation}
    \label{eqn:diagrhotdt}
    \hat{\rho}(t+\dd t)=\sum_{j=1}^{M(t+\dd t)}{\tilde p}_{j}\ketbra*{\tilde{\varphi}_j}\,,
\end{equation}
with $\ket*{\tilde{\varphi}_j}$ the first $M(t+\dd t)$ eigenvectors of $\bm T\bm T^\dagger$ with the largest eigenvalues. The apparent difficulty in diagonalizing the $N_{\mathcal{H}} \times N_{\mathcal{H}}$ density matrix $\bm T\bm T^\dagger$ is avoided by noticing that the $M\times M$ matrix $\bm T^\dagger\bm T$ has the same non-vanishing eigenvalues as its adjoint with the associated eigenvectors being linearly related through $\bm{T}$.

\subsection{First-order equivalence of the two schemes}
\label{sec:equivalence}
In this section we set out to prove the equivalence, to first order in perturbation theory, between the schemes in Refs.~\cite{mccaul2021,chen2021,ciuti2021} and the present LR-TDVP method summarized by the EOM \eqref{eqn:EOM}. 
To do so we make the following two assumptions. First, we assume that at time $t$ the system density matrix $\rhot$ is in its diagonal spectral form \eqref{eqn:diagrhot}, that is: 
\begin{equation}
    \begin{aligned}
        &\B  = \operatorname{diag}\qty(p_1,\ldots,p_M),\\
        &\Bi = \operatorname{diag}\qty(\frac{1}{p_1},\ldots,\frac{1}{p_M}),\\
        &\S  = \Si = \I.
    \end{aligned}    
\end{equation}
Second, we assume the rank $M$ of $\rhot$ to be unchanged at time $t+\dd t$.
Within these assumptions, the variational Eq.~\eqref{eqn:EOM} can be cast in the compact form 
\begin{align}
    \label{eqn:Bdot_diag}
    \Bdot &= \L,\\
    \label{eqn:phidot_diag}
    \ket{\dot\varphi_j} &= \frac{[\I-\P]\LL(\rhoo)}{p_j}\ket{\varphi_j}.
 \end{align}
Note that since the dynamical truncation scheme does not enforce trace preservation, in this section neither do we. Indeed, Eq.~\eqref{eqn:Bdot_diag} does not include the additional term $-\I\Tr{\L}/M$ enforcing trace preservation in Eq.~\eqref{eqn:EOM}. 
The proof now consists in showing that the diagonal matrix \eqref{eqn:diagrhotdt} produced by the dynamical truncation scheme coincides with that generated by Eqs.~\eqref{eqn:Bdot_diag}~and~\eqref{eqn:phidot_diag} to leading order in $\dd t$.
To complete the proof, we express the density operator $\hat{\rho}(t+dt)$ to leading order in $\dd t$ as
\begin{equation}
\begin{aligned}
\label{eqn:rhotdtmat}
\rhoo(t+\dd t)&=\sum_{j=1}^Mp_{j}\ketbra{\varphi_j}+\mathcal{L}[\rhot] \dd t\\
&=\sum_{j=1}^{M}p_{j}\ketbra{\varphi_j}+\sum_{j,k=1}^{N_{\mathcal{H}}}\L_{jk} \ketbra{\varphi_j}{\varphi_k}\dd t\,,
\end{aligned}
\end{equation}
where we have completed the orthogonal basis of the full Hilbert space by introducing a set of orthogonal vectors $\{\ket{\varphi_j},\,j=M+1,\ldots,N_{\mathcal{H}}\}$. The dynamical truncation scheme is achieved by diagonalizing the matrix $\rhoo(t+\dd t)$ in Eq.~(\ref{eqn:rhotdtmat}). The first term in Eq.~\eqref{eqn:rhotdtmat} is diagonal in the LR subspace $\mathcal{H}_M$, and thus has nonzero diagonal elements $p_j$ for $j=1,\ldots,M$ only. The second term is non-diagonal and small. The eigenvalue perturbation theory, therefore, achieves the diagonalization to leading order in $\dd t$. According to perturbation theory, the new eigenvalues and eigenvectors are expressed as
\begin{align}
\label{eqn:pertp}
    \tilde p_j&=p_j+\mathbf{L}_{jj}\dd t\,,\\
    \label{eqn:pertphi}
    \ket{\tilde \varphi_j}&=\ket{\varphi_j}+\sum_{k\ne j}^{1,M}\ket{\varphi_k}\frac{\mathbf{L}_{kj}}{p_j-p_k}\dd t \,\,+ \\
    &\quad+ \qty(\,\sum_{k= M+1}^N\ketbra{\varphi_k})\frac{\mathcal{L}(\hat{\rho})\ket{\varphi_j}}{p_j}\dd t\,,
\end{align}
where we have distinguished the terms with $j=1,\ldots,M$ from the other terms in the sum over states orthogonal to the LR subspace. Notice that in the last term of Eq.~\eqref{eqn:pertphi} the sum over the projectors $\ketbra{\varphi_k}$ factors out, as in the unperturbed matrix $p_k=0$ for $k>M$.
Using that $\I-\P=\sum_{k= M+1}^{N_{\mathcal{H}}}\ketbra{\varphi_k}$, we immediately identify Eq.~\eqref{eqn:phidot_diag} with the last term on the rhs of Eq.~\eqref{eqn:pertphi}. The remaining terms in Eq.~\eqref{eqn:pertphi}, together with Eq.~\eqref{eqn:pertp}, are the perturbation-theory expansion of the $M\times M$ matrix in Eq.~\eqref{eqn:rhotdtmat}. This concludes the proof.

The relation between the truncation error $\epsilon_M$ in Eq.~\eqref{eqn:truncation_error} and the approximated variational error in Eq.~\eqref{eqn:SiL} is also made clear from the first-order perturbative expansion. Indeed, if $\sum_j p_j=1$, then
\begin{equation}
    \epsilon_M = 1-\sum_{j} \tilde p_j = -\Tr{\L}\dd t \propto\Tr{\P\LL(\rhoo)} = \chi\,,
\end{equation}
where we made use of the fact that $\S^{-1} = \I$ for an orthonormal basis.

\end{document}